\documentclass[pdftex,twocolumn,epjc3]{svjour3}          

\RequirePackage[T1]{fontenc}

\smartqed  

\RequirePackage{mathptmx}      
\RequirePackage{flushend}
\RequirePackage[numbers,sort&compress]{natbib}

\usepackage{graphicx,subfig}
\usepackage{float}
\usepackage{bm}
\usepackage{amsmath}
\usepackage{amsfonts,amssymb}
\usepackage{color}
\definecolor{v}{rgb}{0.6, 0.2, 0.8} 
\usepackage{soul} 
\usepackage{enumerate}
\usepackage{float}
\usepackage{subfig}
\usepackage{tabularx}
\usepackage{multicol}        
\usepackage{bm}	
\usepackage[draft]{hyperref}

\journalname{Eur. Phys. J. C}

\begin{document}

\title{Gravitational waves in braneworlds after multi-messenger events}


\author{Miguel A. Garc\'{\i}a-Aspeitia\thanksref{e1,addr1, addr2}
        \and
        Celia Escamilla-Rivera\thanksref{e2,addr3}
}

\thankstext{e1}{e-mail: aspeitia@fisica.uaz.edu.mx}
\thankstext{e2}{e-mail: celia.escamilla@nucleares.unam.mx}

\institute{Unidad Acad\'emica de F\'isica, Universidad Aut\'onoma de Zacatecas, Calzada Solidaridad esquina con Paseo a la Bufa S/N C.P. 98060, Zacatecas, M\'exico \label{addr1}
          \and
          Consejo Nacional de Ciencia y Tecnolog\'ia, Av. Insurgentes Sur 1582. Colonia Cr\'edito Constructor, Del. Benito Ju\'arez C.P. 03940, Ciudad de M\'exico, M\'exico. \label{addr2}
          \and
          Instituto de Ciencias Nucleares, Universidad Nacional Aut\'onoma de M\'exico, Circuito Exterior C.U., A.P. 70-543, M\'exico D.F. 04510, M\'exico. \label{addr3}
}

\date{Received: date / Accepted: date}

\maketitle

\begin{abstract}
The GW170817 event opened a new window to test modifications to General Relativity with the aim of discard or impose strong constraints in extra dimension theories of gravity. Concerning these theories, the Randall-Sundrum brane-world theory -- 4+1 spacetime model where its covariant Einstein field equations are composed with new extra terms that comes from extra dimensions -- is begin tested in the multi-messenger astronomy field. In the same context, the aim of this paper is to impose new constrictions through measurements of the time-delay between the gravitational and electromagnetic signals: GW170817 and the recent S190425z, over the free cosmological parameters of this modified model. We assume that Gravitational Waves travel along a shortcut on the extra dimension. In addition, we consider the standard $\Lambda$CDM model and perform a likelihood analysis in order to study effects like the $H_0$  tension in comparison to this extra-dimensional theory, obtaining a reduced $H_0$ value of the order of $0.2\%$ in the low energy limit of this theory.
\end{abstract}



\section{Introduction} \label{Int}

Currently, the GW170817 event \cite{Abbott} and the recent S190425z detection  \cite{Pozanenko:2019lwh}, related with the multi-messenger Gravitational Waves (GWs) astronomy and subsequently electromagnetic counterparts \cite{AbbottII}, have been useful to set constraints on viable gravitational theories. Being an interesting way to explore gravity with potential new applications to test the astrophysics of binary neutron stars merger \cite{Faber}, cosmic expansion \cite{Nissanke:2013fka} and even modifications to the General Theory of Relativity (GR) (see for instance \cite{Visinelli:2017bny,Gumrukcuoglu:2017ijh,Dima:2017pwp,Gong:2017kim,Ezquiaga:2017ekz}) and its consequences.

In particular, theories that modify gravity have suffered a severe setback, due the constraint imposed by these astrophysical events, being some of them, the scalar-tensor theories and others dark energy (DE) models that predict a different speed of propagation for GW (see \cite{Visinelli:2017bny,Gumrukcuoglu:2017ijh,Dima:2017pwp,Gong:2017kim,Ezquiaga:2017ekz,PhysRevD.79.064036,Deffayet:2009wt,Charmousis:2011bf} for details). Hence, several proposals have been focused in study the viability of models like Horndeski \cite{Horn}, beyond Horndeski \cite{Zumalacarregui:2013pma}, degenerated higher-order-scalar-tensor \cite{Crisostomi:2016czh,Achour:2016rkg}, among others, discarding some of them or establishing strong constraints in their free parameters.

Observations from the GW that came from the binary neutron stars merger, which was detected by the LIGO-VIRGO collaboration, and the one followed-up by a short gamma ray burst (GRB170817A), seeing just $1.75\pm0.05$, later by Fermi and the International Gamma-Ray Astrophysical Laboratory \cite{AbbottII}, have been used  in order to understand the constraints over the cosmological parameters in the theory. Now, with these observations we can set a physical range of $[0.86-2.26]$M$_\odot$ to weight the neutron star and to measure the luminosity distance $d_L=40^{+8}_{-14}$Mpc, which is a direct result from the bound $-3\times10^{-15}\leqslant c_g/c-1\leqslant7\times10^{-16}$, which was found under the lowest limit for $d_L$ and the corrective delay between the GW and the gamma ray burst\footnote{Following the recipe of \cite{Ezquiaga:2017ekz}, we will consider a cutoff of $\vert c_g-1\vert\leqslant 5\times10^{-16}$.}. Moreover, the most recent detection of GWs are given by the event so-called S190425z registered by LIGO-VIRGO detectors \cite{Pozanenko:2019lwh}. This event is a high confidence ($>99\%$) neutron stars merger with an electromagnetic counterpart caused by a weak gamma ray burst (GRB190425), being the only second event known in addition to GW170817. Both events, provide higher constrictions on GR or even for extensions or modifications to this fundamental theory.

From a theoretical point of view, brane world theories is an alternatively form to understand the nature of gravity and a feasible direction to understand puzzles like the dark matter and dark energy problems. In this regards, the Randall-Sundrum models (RS) and its extensions are of great interest in the community  \cite{Randall-I,Randall-II}. The idea of RS branes is based in the proposition that the universe is conformed by a 4+1 dimensional space-time called the bulk, which is usually assumed anti-deSitter (AdS$_5$) and where it is immersed a four dimensional manifold called the brane with a metric that depends of what we are studying. The existence of extra dimensions in this framework and in particular, the modifications produced by braneworlds, can be constrained under the free parameter $\lambda$, which is known as the \textit{brane tension}. This latter being the parameter that relates the four dimensional structure with the five dimensional bulk. Several proposals have constrained the brane tension parameter as $\lambda\simeq6.42h^2\times10^5$eV$^4$ using a joint analysis \cite{Garcia-Aspeitia:2016kak} of cosmological observations or $\lambda\simeq138.59\times10^{48}$eV$^4$ through Table-Top experiments \cite{Gergely} at astrophysical level. The results derived from them are considerable promising, constraining the brane tension of the order $\lambda\simeq10^{32}$eV$^4$ \cite{Germani:2001du,Garcia-Aspeitia:2015mwa,Linares:2015fsa}. Natural extensions to brane world theories consider extra degree of freedom in the brane tension, which now depend of the time variable \cite{Guendelman:2002mf,HoffdaSilva,Gergely}, showing a better concordance with cosmological constraints (see as e.g  \cite{Garcia-Aspeitia:2018fvw}). 

On the other hand, some attempts have been done in order to set a cutoff on the screening scale using gravitational and electromagnetic waves, e.g for theories beyond 4 dimensions \cite{Pardo:2018ipy}, where 20 Mpc scales needs to be consider. Also, the addition of EM counterparts as the GRB170817A \cite{Monitor:2017mdv} was used in \cite{Visinelli:2017bny} to determine an upper bound of the radius of curvature.

Therefore, it is reasonable to propose new constraints in order to validate or discard theories based on the RS hypothesis. Hence, GW observations provide important constraints for RS theories and specifically after GW170817 and S190425z for the propagation speed in vacuum. Is with this idea, that we propose extra constrictions through GWs, both at the low and high field level as a complement of the previous studies presented in the literature. 

The aim of this paper is to constraint RS branes using the covariant form of the modified Einstein field equations \cite{sms} in the weak field limit. In this scenario, the vacuum solution for the wave equation is not the traditional due that extra terms that come from Weyl terms could play an important role in the propagation and speed of GWs. Previous results have been focused in weak field limit, studying the consequences of the quadratic part of the energy-momentum tensor in the dynamics of binary system (see for instance \cite{Garcia-Aspeitia:2013jea}) and also in extended theories of gravity \cite{Nunes:2018evm,Escamilla-Rivera:2019ulu}.

In order to study the low (Minkowski) and high (De Sitter) energy limits of the extra-dimensional theory, we compute the exact solutions for both limits and adopt a Bayesian approach to combine two GW detections. We set the free cosmological parameters $\theta$ taking into account the $\Lambda$CDM as pivot model. Via Bayes theorem, the probability is $p\propto (\theta, \kappa, d_{L})$, where the former if the joint likelihood and the latter if the cosmological prior. As was pointed out in \cite{Visinelli:2017bny}, we can consider an upper limit of $\ell < 0.535$ Mpc in order to study the $H_0$ tension at low and high energy limits in this model\footnote{Recently in Ref. \cite{Lin:2020wnp} the authors constraint $\ell\gtrsim7.5\times10^2$Tpc.}.

The outline of the paper is as follows: in Sec. \ref{sec:WF}, we show the covariant form of the modified Einstein equations, identifying the source and geometrical terms in the field equations in vacuum. We study these equations in the weak field limit, showing that the canonical wave equation contains non-local terms that come from branes. Our results are compatible with the expected constrictions about the speed propagation of gravitons and it is constrained with other observations. In Sec. \ref{Sec:Perturbations} we  discuss the brane world cosmic perturbations, focusing our attention in the Minkowski and De Sitter brane cases. Sec. \ref{Sec:Limits} is dedicated to set cosmological constraints with the braneworld GW propagation through the events GW170817 and S190425Z. Finally, discussions and conclusions are presented in Sec. \ref{DC}. Throughout the text, we will use natural units, unless we indicate otherwise.

\section{Braneworld gravitational waves in weak field limit} 
\label{sec:WF}

The field equations without the presence of matter fields in the bulk can be written as $G_{\mu\nu}+\xi_{\mu\nu}=\kappa^2_{(4)}S_{\mu\nu}$, where $S_{\mu\nu}=T_{\mu\nu}+6\Pi_{\mu\nu}/\lambda$. Here the l.h.s of the field equation contains the geometric part with $G_{\mu\nu}$ as the Einstein tensor and $\xi_{\mu\nu}$ related with the non-local Weyl tensor that comes from branes contributions. Hence, the r.h.s of this field equation contains the source of matter and energy with $T_{\mu\nu}$ as the energy-momentum tensor and $\Pi_{\mu\nu}$ the quadratic part of the energy-momentum tensor. In addition, $\kappa^2_{(4)}\equiv 8\pi G$ is the four dimensional coupling constant, $G$ is the gravitational Newton constant and $\lambda$ is the brane tension parameter, which is the threshold between the traditional Einstein field equations and the corrections caused by the presence of branes. Notice that the brane tension helps to quantify the presence of extra dimensions in various phenomena studied. 
\\

In this landscape we define
\begin{eqnarray}
\xi_{\mu\nu}&\equiv&C_{AEFB}n^En^Bg^A_{\mu}g^F_{\nu}, \label{C}\\
\Pi_{\mu\nu}&\equiv&-\frac{1}{4}T_{\mu\alpha}T^{\alpha}_{\nu}+\frac{1}{12}T^{\alpha}_{\alpha}T_{\mu\nu}+\frac{1}{24}g_{\mu\nu}(3T_{\alpha\beta}T^{\alpha\beta} 
-T^2), \quad\quad \label{Pi}
\end{eqnarray}
where $C_{AEFB}$ is the five dimensional Weyl tensor, $n^{E}$ is a unitary normal vector and $T\equiv T^{\alpha}_{\alpha}$.
The symmetry of $\xi_{\mu\nu}$ imply that we can decompose it irreducibility as
\begin{eqnarray}
\xi_{\mu\nu}&=&-\frac{6}{\kappa^2_{(4)}\lambda}\left[\mathcal{U}\left(u_{\mu}u_{\nu}+\frac{1}{3}\epsilon_{\mu\nu}\right)+\mathcal{P}_{\mu\nu}\right],
\end{eqnarray}
where $\mathcal{U}$ is the non-local energy density and $\mathcal{P}_{\mu\nu}$ is the non-local anisotropic stress tensor, $u_{\alpha}$ is the four-speed, which also satisfy $g_{\mu\nu}u^{\mu}u^{\nu}=-1$, and $\epsilon_{\mu\nu}=g_{\mu\nu}+u_{\mu}u_{\nu}$ is orthogonal to $u_{\mu}$. Notice that
\begin{equation}
\mathcal{U}\equiv-\frac{\kappa^2_{(4)}\lambda}{6}\xi_{\mu\nu}u^{\mu}u^{\nu}, \;\; \mathcal{P}_{\mu\nu}\equiv-\frac{\kappa^2_{(4)}\lambda}{6}\xi_{\langle\mu\nu\rangle}.
\end{equation}

We start the analysis in the weak-field limit imposing $g_{\mu\nu}=\eta_{\mu\nu}+h_{\mu\nu}$, where $\eta_{\mu\nu}$ is the Minkowski metric, $\eta_{\mu\nu}={\rm diag}(-1,1,1,1)$ and $\vert h_{\mu\nu}\vert\ll1$ is a small perturbation. Therefore, in Einstein gauge (harmonic coordinates)\footnote{We set the Einstein gauge in the form $\tilde{h}^{\mu}_{\nu,\mu}=h^{\mu}_{\nu,\mu}-\frac{1}{2}h_{,\nu}=0$ where $\tilde{h}_{\mu\nu}=h_{\mu\nu}-\frac{1}{2}\eta_{\mu\nu}h$.}, the field equations reduce to
\begin{eqnarray}
&&\Box h_{\mu\nu}-\frac{2}{\kappa^2_{(4)}\lambda}\left[\mathcal{U}\left(4u_{\mu}u_{\nu}+\eta_{\mu\nu}\right)+3\mathcal{P}_{\mu\nu}\right]=
\kappa^2_{(4)}S^{(0)}_{\mu\nu}, \label{field}
\end{eqnarray}
and in the weak field limit the non-local term can be reduced to
\begin{eqnarray}
\xi_{\mu\nu}^{(0)}&=& \left[2h_{\alpha\delta,\beta\gamma}-\eta_{\alpha[\gamma}\Box h_{\delta]\beta}+\eta_{\beta[\gamma}\Box h_{\delta]\alpha}\right]
n^{\beta}n^{\delta}\eta^{\alpha}_{\mu}\eta^{\gamma}_{\nu}\nonumber\\&&+\mathcal{O}(5D), \label{xi}
\end{eqnarray}
where the d'Alambertian operator is $\Box\equiv\eta^{\mu\nu}\partial_{\mu}\partial_{\nu}$. In this case, we only consider the pure four dimensional part of the Weyl tensor, i.e. $C_{AEFB}\to C_{\mu\nu\gamma\beta}$  (combinations with four and five dimensions and pure five dimensional terms are encoded in $\mathcal{O}(5D)$), under the assumption that is the only part that can be detected by LIGO and VIRGO interferometers. We also have that $S_{\mu\nu}$ and $\xi_{\mu\nu}$ are taken to lowest order in $h_{\mu\nu}$, denoted with the superscript $(0)$. Due that the Weyl tensor is traceless, $\Box h=0$, the $S_{\mu\nu}$ tensor at lowest order will only contain contributions to the traditional energy-momentum tensor $T_{\mu\nu}$, because it will have negligible contributions caused by $\Pi_{\mu\nu}$ since it contains contributions at higher order in the energy-momentum tensor as it is shown in Eq. \eqref{Pi}, therefore $S_{\mu\nu}^{(0)}\to M_{\mu\nu}^{(0)}\equiv T_{\mu\nu}-\frac{1}{2}\eta_{\mu\nu}T^{\lambda}_{\lambda}$.

In vacuum, the field equations do not depend of the energy-momentum tensor (and quadratic part) source i.e. $M^{(0)}_{\mu\nu}=0$. However, it depends of the non-local geometric contribution encoded in $\xi_{\mu\nu}^{(0)}$. Notice how the propagation of GWs depend of the presence of the r.h.s term shown in Eq. \eqref{Vac} in the form
\begin{eqnarray}
&&\Box h_{\mu\nu}=\frac{2}{\kappa^2_{(4)}\lambda}\left[\mathcal{U}\left(4u_{\mu}u_{\nu}+\eta_{\mu\nu}\right)+3\mathcal{P}_{\mu\nu}\right]=-\xi_{\mu\nu}^{(0)}. \label{Vac}
\end{eqnarray}
We can rewrite the later using Eq. \eqref{xi} to obtain
\begin{eqnarray}
&&\Box h_{\mu\nu}-(\eta_{\alpha[\gamma}\Box h_{\delta]\beta}-\eta_{\beta[\gamma}\Box h_{\delta]\alpha}-2h_{\alpha\delta,\beta\gamma})n^{\beta}n^{\delta}\eta^{\alpha}_{\mu}\eta^{\gamma}_{\nu}
\nonumber\\&&
+\mathcal{O}(5D)=0. \label{reac}
\end{eqnarray}
The l.h.s contains extra terms in comparison to the standard GR and giving extra information provided by the Weyl tensor. Each term in Eq. \eqref{reac} only has a geometric nature, remarking only the 4D characteristics, while 5D contributions are hidden from direct observations. Information about the propagation speed is encoded in Eq. \eqref{reac}, being able to be constrained by the GWs observational events.

In order to extract information of Eq.\eqref{reac}, we adopt the ansatz $h_{\mu\nu}=h_{\mu\nu}(t-x)$. At this point it is possible to infer the only two non-negligible functions as: $h_{22}=h_{22}(t-x)$ and $h_{23}=h_{23}(t-x)$, related with the polarization states. Notice that $h_{\mu\nu}(t-x)$ is equal to zero. Therefore, Eq. \eqref{reac} has extra terms only when $\mu=\delta=2$ and $\nu=\beta=3$, resulting in 
\begin{equation}
(\partial^2_x-\partial^2_t)(h_{23}+ h_{22}n^3n^2)+\mathcal{O}(5D)=0.
\end{equation}
More cases are reported in \ref{Appendix}.

As we expect, normal unitary vectors are orthogonal to each other and we recover the form  $(\partial^2_x-\partial^2_t) h_{23}+\mathcal{O}(5D)=0$. The other case is $(\partial^2_x-\partial^2_t) h_{22}+\mathcal{O}(5D)=0$. Hence we obtain the standard wave equation for a GW propagating at the speed of light and cannot be ruled  out by the GWs events. Additional information comes from 5D contributions and also has dependence of the coupling constant $\lambda$, which can be constrained. We notice that this parameter cannot be detected by the LIGO antenna. 
\\

The l.h.s  has the d'Alambertian operator associated to the perturbation and the r.h.s represent the double derivative of the perturbation $h_{\alpha\beta}$, which is associated with the Riemann tensor. 
Here the propagation speed must be equal to the speed of light as in GR, coinciding with the GW170817 event. \\

Furthermore, notice that the solution of Eq. \eqref{Vac}, must be the retarded potential written as
\begin{eqnarray}
h_{\mu\nu}({\bf x},t)&=&-\frac{1}{2\pi\kappa^2_{(4)}\lambda}\int\left[4\mathcal{U}u_{\mu}u_{\nu}+\mathcal{U}\eta_{\mu\nu}+3\mathcal{P}_{\mu\nu}\right]\times\nonumber\\&&\frac{({\bf x^{\prime}},t-\vert {\bf x}-{\bf x^{\prime}}\vert)}{\vert {\bf x}-{\bf x^{\prime}}\vert}d^3{\bf x^{\prime}}, \label{1GW}
\end{eqnarray}
or in general with the presence of a energy momentum tensor
\begin{eqnarray}
h_{\mu\nu}({\bf x},t)&=&\frac{\kappa^2_{(4)}}{2\pi}\int\frac{M^{(0)}_{\mu\nu}({\bf x^{\prime}},t-\vert {\bf x}-{\bf x^{\prime}}\vert)}{\vert {\bf x}-{\bf x^{\prime}}\vert}d^3{\bf x^{\prime}}\nonumber\\&&-\frac{1}{2\pi\kappa^2_{(4)}\lambda}\int\left[4\mathcal{U}u_{\mu}u_{\nu}+\mathcal{U}\eta_{\mu\nu}+3\mathcal{P}_{\mu\nu}\right]\times\nonumber\\&&\frac{({\bf x^{\prime}},t-\vert {\bf x}-{\bf x^{\prime}}\vert)}{\vert {\bf x}-{\bf x^{\prime}}\vert}d^3{\bf x^{\prime}}. \label{01GW}
\end{eqnarray}
This result represent GWs in this framework that not only can be produced by the presence of the energy-momentum tensor, but also by non-local effects given by the Weyl tensor \eqref{1GW} or from \eqref{reac}. Therefore, we can write
\begin{eqnarray}
&&h_{\mu\nu}({\bf x},t)=\frac{1}{2\pi}\int\Big[h_{\alpha\delta}({\bf x},t)_{,\beta\gamma}-\frac{1}{4}(\eta_{\alpha[\gamma}\Box h_{\delta]\beta}({\bf x},t)\nonumber\\&&-\eta_{\beta[\gamma}\Box h_{\delta]\alpha}({\bf x},t))\Big]\times\frac{n^{\beta}n^{\delta}\eta^{\alpha}_{\mu}\eta^{\gamma}_{\nu}({\bf x^{\prime}},t-\vert {\bf x}-{\bf x^{\prime}}\vert)}{\vert {\bf x}-{\bf x^{\prime}}\vert}d^3{\bf x^{\prime}}\nonumber\\&&+\mathcal{O}(5D). \label{2GW}
\end{eqnarray}
In vacuum and under the Bianchi identities on the brane $\xi^{\mu}_{\nu,\mu}=0$, the Einstein gauge is fulfilled $\xi^{\mu}_{\nu,\mu}-\xi_{,\nu}/2=0$ for a traceless Weyl tensor. 
The extra term in Eq. \eqref{01GW} can be constrained, e.g with Astrophysics \cite{Germani:2001du,Linares:2015fsa} or Table Top experiments \cite{Gergely_2006}, giving a brane tension of the order $\lambda=5\times10^{32}$eV$^4$ and $\lambda=138.59\times10^{48}$eV$^4$, respectively. These correction are of the order $\sim2.88\times10^{23}$eV$^{-2}$ and $\sim1.03\times10^6$eV$^{-2}$, for astrophysics and table top experiments respectively, for the extra-dimensions contributions.  

\section{Braneworld cosmic tensor perturbations}
\label{Sec:Perturbations}

Let us consider bulk Einstein equations $R_{\mu\nu}=0$ (with $\Lambda=0$). By Gauss relation we obtain the Ricci tensor
\begin{equation}\label{eq:Runperturbed}
\bar{R}_{\mu\nu}= K^{\alpha}_{\alpha}K_{\mu\nu} -K^{\rho}_{\mu} K_{\nu\rho} -\bar{\xi}_{\mu\nu},
\end{equation}
where $K_{\mu\nu}= -\bar{g}^{\rho}_{\mu}\nabla_{\rho} n_{\nu}$ is the extrinsic curvature tensor of the brane. $\bar{\xi}_{\mu\nu}\equiv C_{\mu\rho\nu\sigma} n^{\rho}n^{\sigma}$ is the contribution from Weyl, where the bar denote curvature tensors pertaining to the brane metric $\bar{g}_{\alpha\beta}$. Using the junction conditions \cite{gravitation,Tavakoli:2015llr} we obtain for the extrinsic curvature
\begin{equation}
K_{\alpha\beta} = \frac{4\pi G}{3} \left[3T_{\alpha\beta} +(T^{\alpha}_{\alpha} -\lambda) \bar{g}_{\alpha\beta}\right].
\end{equation}
Combining with the bulk Einstein equations we obtain the Lie derivatives in the normal direction to the brane, which we need to linearize 
\begin{subequations}
\begin{eqnarray}
D_{n} K_{\alpha\beta} &=& \bar{\xi}_{\alpha\beta} -K_{\alpha\beta} K^{\gamma}_{\beta}, \\
D_{n} \bar{g}_{\alpha\beta} &=& -2 K_{\alpha\beta}. 
\end{eqnarray} \label{eq:lie}
\end{subequations}

We consider now a background metric
\begin{equation} \label{eq:backm}
ds^2 = -n^2 (\tau,\chi) d\tau^2 +a^2 (\tau,\chi) \delta_{ij} dx^{i}dx^{j} +d\chi^2,
\end{equation}
where $n$ and $a$ are functions to be found and it is also considered that the fifth dimension is static. 
Therefore, we can write the perturbed metric for AdS bulk  in the form
\begin{equation}
ds^2 = \left(\frac{\ell}{z}\right)^2 (\eta_{\alpha\beta} +h_{\alpha\beta}) dx^\alpha dx^\beta,
\end{equation}
where $z=\ell/a$, set a new place of the brane, $\eta_{\alpha\beta}$ is the Minkowski metric tensor and $h_{\alpha\beta}\ll1$ is a small perturbation to the Minkowski background metric. In addition, we choose the transverse-traceless (TT) gauge with $\eta^{\alpha\beta}h_{\alpha\beta}=0$ and $\partial_{\alpha} h^{\alpha\beta}=0$. We require that the perturbed metric have the form
\begin{equation}
ds^2 = -n^2 (\tau,\chi)d\tau^2+a^2(\tau,\chi)[\delta_{ij}+h_{ij}]dx^i dx^j +d\chi^2.
\end{equation}
The previous metric only contains the tensor perturbations, neglecting scalar and vector perturbations. \\

For the energy momentum tensor background we have $T^{\alpha}_\beta =\text{diag} (-\rho,p\delta_{ij})$, and for its perturbation $\delta T^0_0=-\delta\rho$, $\delta T^0_i=\delta P_i$, $\delta T^i_j=\delta p\delta^i_j+\delta\pi^i_j$, where $\delta\rho$ and $\delta p_i$ are perturbed density and pressure respectively, $\delta P_i$ is the perturbed three momentum and $\delta\pi^i_j$ is the traceless anisotropic stress. \\

The variation of the spatial part is $\delta \bar{R}_{ij} =\delta R^{0}_{i0j} +\delta \bar{R}^s_{isj}$ and for the tensor part:
\begin{eqnarray}\label{eq:pertR1}
\delta \bar{R}_{ij}&=& \frac{a^2}{2n^2} \dot{h}_{ij} + \frac{a^2}{2n^2}\left(3\frac{\dot{a}}{a}-\frac{\dot{n}}{n}\right)h_{ij}  
\nonumber\\&&
+\frac{a^2}{n^2}
\left(\frac{\ddot{a}}{a} -\frac{\dot{a}\dot{n}}{an}+2\frac{\dot{a}^2}{a^2}\right)h_{ij} +\frac{1}{2}k^2 h_{ij},
\end{eqnarray}
where the dots represents derivatives with respect to the coordinate time $\tau$. The mode $k$ comes from the decomposition in Fourier modes of the perturbed metric as $h^{i}_{j}(\tau,\chi;x^l) = h(\tau,\chi;k^l) e^{ik^l\cdot x^l}\hat{e}^{i}_{j}$, where $\hat{e}^{i}_{j}$ is the trace-transverse polarization, $\nabla^2\hat{e}_{ij}=-k^2\hat{e}_{ij}$.

Rewriting the perturbation of the Ricci tensor in the form:
\begin{equation}\label{eq:pertR2}
\delta \bar{R}_{ij} = 2{a^\prime}^{2}h_{ij} +\frac{1}{2}\left(\frac{n^\prime}{n}-\frac{a^{\prime}}{a}\right) \frac{d}{d\chi}(a^2 h_{ij}) +\frac{1}{2}\frac{d^2}{d\chi^2} (a^2 h_{ij}),
\end{equation}
where the primes represents derivatives with respect to the $\chi$ parameter and the spatial part of the Ricci tensor from the background metric takes the form
\begin{eqnarray}
\frac{\ddot{a}}{an^2} +2\frac{\dot{a}^2}{a^2n^2} -\frac{\dot{a}\dot{n}}{an^3} -\frac{a^{\prime\prime}}{a} -\frac{a^\prime n^\prime}{an} -2\frac{{a^\prime}^2}{a^2}=0.
\end{eqnarray}
From (\ref{eq:pertR1}) and (\ref{eq:pertR2}) and using the latter we obtain the RS graviton equation in vacuum 
\begin{eqnarray}\label{eq:RSgraviton}
&&\ddot{h}_{ij} + \left(3\frac{\dot{a}}{a} -\frac{\dot{n}}{n}\right)\dot{h}_{ij} +k^2 \left(\frac{n^2}{a^2}\right) h_{ij} -n^2 \left(3\frac{a^\prime}{a}+\frac{n^\prime}{n}\right)h_{ij}^{\prime} 
\nonumber\\&& 
-n^2 h_{ij}^{\prime\prime} =0,
\end{eqnarray}
where $h_{ij} = h^{+}e^{+}_{ij} +h^{\times}e^{\times}_{ij}$ with two polarization tensors $e^{\times,+}_{ij}$. From the later, we can related the sound speed of the graviton $c_{s}^2 =n^2$. The first GW detection impose $c_{s}^2\approx 1$ which implies $n\approx1$.  \\

Conversely, we can associate the friction term from the above gravitational wave equation which can denote how the GW amplitude decreases in the propagation across cosmological distances, from the source to the observer. In particular,  for inspiraling binaries this factor combines with other factors coming from the transformation of masses and frequency from the source frame to the detector frame. To produce  the usual dependence of the GW amplitude we write the luminosity distance as
\begin{equation}
\tilde{h}_{ij}(\eta, \vec{k})\propto \frac{1}{d_L(z)}\, ,
\end{equation}
this friction term would modify the propagation speed of GWs and here $\eta$ represents the conformal time related with $\tau$. We can now excluded this coefficient 
at the level  $|c_{\rm gw}-c|/c< O(10^{-15})$, by the observation of GW170817 and S190-425z is a likely binary neutron star (BNS) merger at $d_{L}=156\pm 41$ Mpc. These kind of constrains have ruled out a large class of scalar-tensor and vector-tensor modifications of GR. To eliminate the friction term, we introduce $\tilde{\eta}_{ij}(\eta, \vec{k})$ from 
\begin{equation}
\tilde{h}_{ij}(\eta, \vec{k})=\frac{1}{\tilde{a}(\eta)}  \tilde{\eta}_{ij}(\eta, \vec{k})\, ,
\end{equation}
where 
\begin{equation}
\frac{1}{\tilde{a}}\frac{d\tilde{a}}{d\eta}={\cal H}[1-\delta(\eta)]\,.
\end{equation}
We perform the change to conformal time with $\delta(\eta)$ a function that parameterize the deviation from GR. Integrating the latter we obtain
\begin{equation}
d_L^{\,\rm gw}(z)=d_L^{\,\rm em}(z)\exp\left\{-\int_0^z \,\frac{dz'}{1+z'}\,\delta(z')\right\}\,. \label{eq:dl_gw}
\end{equation}
Notice that we must distinguish between the usual 
luminosity distance appropriate for electromagnetic signal, $d_L^{\,\rm em}(z)$ and a GW \textit{luminosity} distance $d_L^{\,\rm gw}(z)$. Based in these definitions, we set the scenario in where we can analyze low (Minkowski) and high (De Sitter) energy regime considering the following two exact solutions.

\subsection{Minkowski brane case}

To analyze the low energy regime of Eq.(\ref{eq:RSgraviton}) we propose a solution given by:
\begin{equation}
a(\tau,\chi) = n(\tau,\chi) =\exp(-\chi/\ell),
\end{equation}
where $\ell$ is the curvature radius of AdS$_5$ (or also known as the length scale). According to this proposal, the axis normal to the brane is constrained through GW170817 with  $\chi \approx 0$. Therefore, we obtain the wave equation: 
\begin{eqnarray}
\ddot{h} +k^2 h = e^{-2\chi/\ell} \left(h^{\prime\prime} -\frac{4}{\ell}h^{\prime}\right). \label{eq:Minkowski brane}
\end{eqnarray}  
Considering separation of variables by multiplying two functions as $h(\tau,\chi)=A(\tau)C(\chi)$, we perform the second derivations with respect to $\tau$ and $\chi$, respectively, therefore we have the following pair of wave evolution equations at l.h.s and with their solutions at r.h.s
\begin{eqnarray}
&&\ddot{A} +(k^2 +m^2) A=0, \rightarrow  A\propto e^{i\sqrt{k^2+m^2}\tau}, \\
&&C^{\prime\prime} -4\ell^{-1}C^{\prime} +m^2 e^{2\chi/\ell}C= 0,  \rightarrow  C\propto e^{2\chi/\ell}J(2,m\ell \sqrt{e^{2\chi/\ell}}). \quad
\end{eqnarray}
Notice that with this proposed solution we decouple the dependence between $\tau$ and $\chi$ and recover a system with exponential behavior depending of the mode $k$ and an integration constant $m$. Also the second solution is weighted by a $J(\alpha,\beta)$ which is the Bessel function of the first kind. 
\\

\begin{figure}
\centering
\includegraphics[width=0.5\textwidth,origin=c,angle=0]{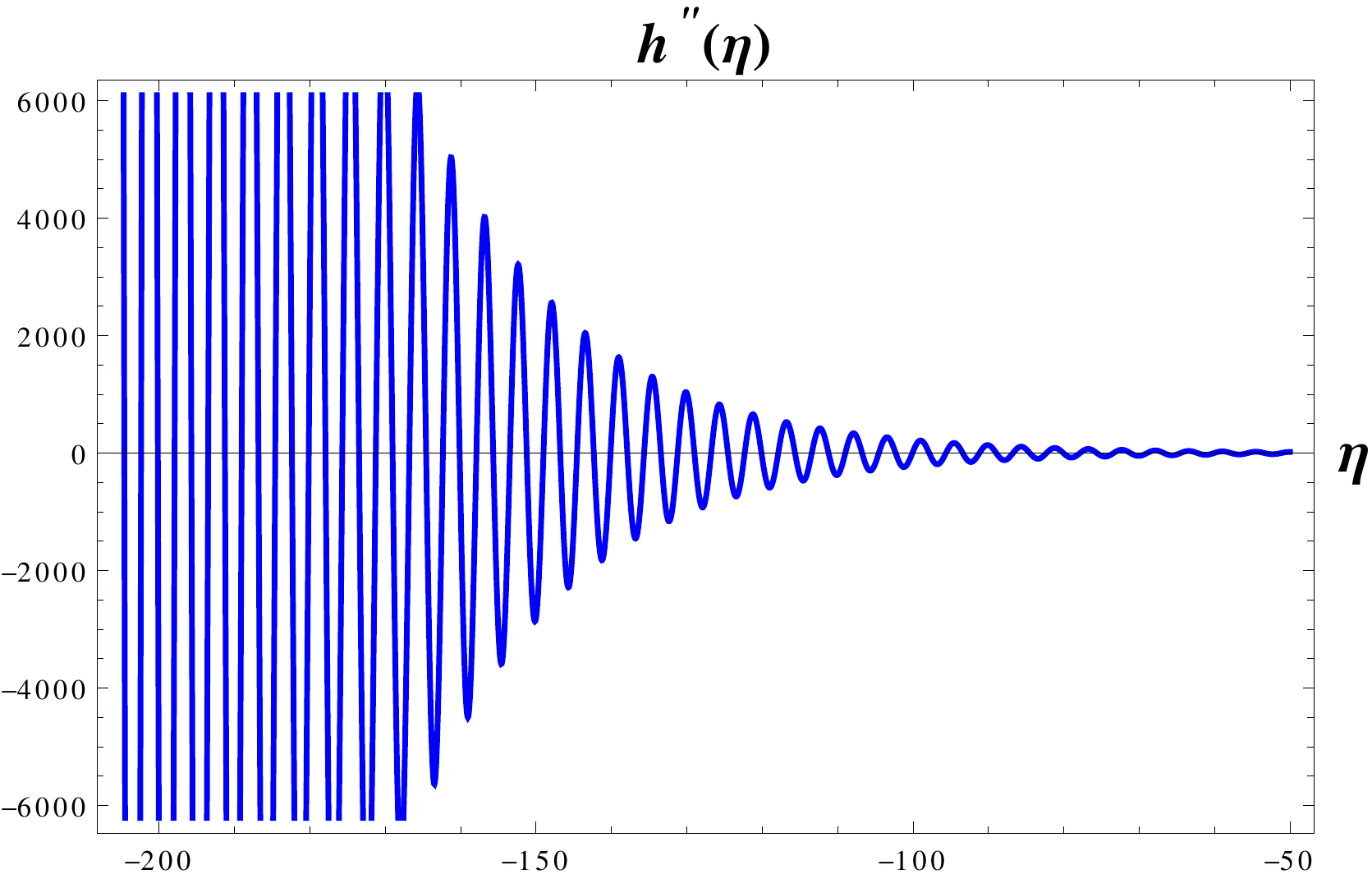}
\caption{ Numerical solution for the evolution of the GW for the Minkowski brane case Eq. \eqref{eq:Minkowski brane} (only temporal dependence) in terms of the conformal time $\eta$, where we used the notation $h''(\eta)\equiv d^2 h/d\eta^2$. The numerical solution is running with values for $k=0.01 \text{s}^{-1}$ and $\ell\gtrsim7.5\times10^2$Tpc \cite{Lin:2020wnp}.} 
\label{evolution_Minkowski}
\end{figure}

In addition, from the numerical solution of Eq. \eqref{eq:Minkowski brane} which it is shown in Figure \ref{evolution_Minkowski}, we notice that function decay faster at large conformal times. The free parameters $\ell$ and $k$ are elected through recent constraints presented in Ref. \cite{Lin:2020wnp}.

\subsection{De Sitter brane case}

To analyze the high energy regime of Eq.(\ref{eq:RSgraviton}) we propose a solution assuming the condition for a constant density\footnote{Recent paper is posted on the pre-print repository, studying the De sitter case \cite{Lin:2020wnp}, constraining the curvature radius as $\ell\gtrsim7.5\times10^2$Tpc. for a constant $\rho_0$.} $\rho_{\text{matter}}\approx \rho_0$. Therefore, we consider a solution given by
\begin{equation}
a(\tau,\chi) = n(\tau,\chi) =a_0 (\tau) F(\chi),
\end{equation}
 where
\begin{equation}
F(\chi) = e^{-\chi/\ell}+\frac{\rho}{2\lambda}( e^{\chi/\ell} -e^{-\chi/\ell}).
\end{equation}
Again, to constraint through GW170817 event we require $F(\chi)\approx 1$, therefore $\rho\approx 2\lambda(1+e^{\chi/\ell})^{-1}$. Using the latter proposed solution we rewrite Eq. (\ref{eq:RSgraviton}) to obtain
\begin{eqnarray}
\ddot{h}+2\frac{\dot{a_0}}{a_0} \dot{h} +k^2 h =a_0^2F^2h^{\prime\prime}+4a_0^2FF^{\prime}h^{\prime}. \label{eq:desitter brane}
\end{eqnarray}
Using separation of variables to decouple both $\tau$ and $\chi$, $h(\tau,\chi)=A(\tau)C(\chi)$, it is straightforward to compute the following differential equations at l.h.s with their respectively solutions at r.h.s
\begin{eqnarray}
&&\ddot{A} +2\frac{\dot{a}_0}{a_0} \dot{A}+(k^2 +m^2 {a_0}^2) A=0,  \rightarrow A\propto \tau^{3/2\pm \alpha},\quad \\ 
&&C^{\prime\prime} +4\frac{F^\prime}{F}C^{\prime} +\frac{m^2}{F^2} C= 0, \rightarrow  C\propto e^{2\chi/\ell}J(2,m\ell \sqrt{e^{2\chi/\ell}}). \quad\quad
\end{eqnarray}
In Eq. (35) we assume $\rho/2\lambda\ll1$, implying $F(\chi)\approx e^{-\chi/\ell}$, in order to have an analytic continuation, being $J(\alpha,\beta)$ the Bessel function of the first kind.
\\

\begin{figure*}
\centering
\includegraphics[width=0.5\textwidth,origin=c,angle=0]{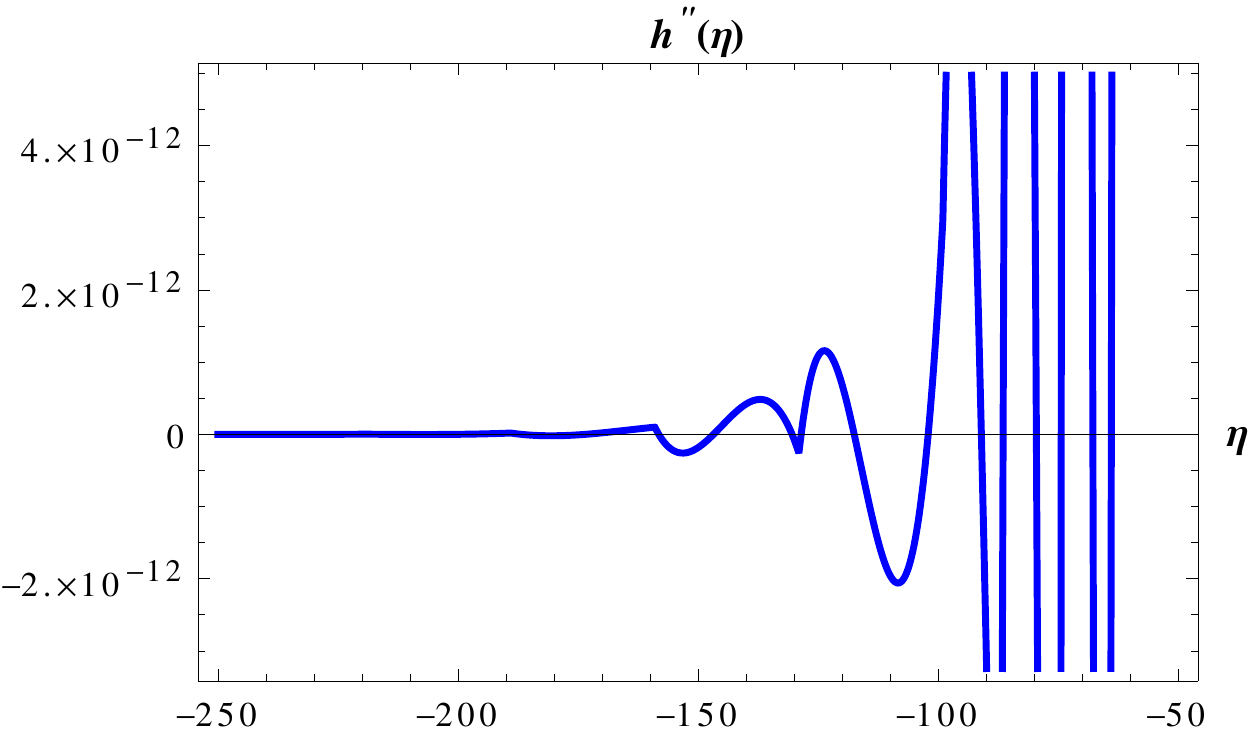}
\caption{Numerical solution for the evolution of the GWs for the De Sitter brane case Eq. \eqref{eq:desitter brane} in terms of the conformal time $\eta$, where we used the notation $h''(\eta)\equiv d^2 h/d\eta^2$. The numerical solution is running with values for $k=0.01\text{s}^{-1}$, $\ell\gtrsim7.5\times10^2$Tpc. We consider also $\Omega_{GW} \sim 10^{-5}$\cite{Turner:1996ck}, an estimate of a flat spectrum across LIGO's band as $\Omega_{GW}=\rho^{-1}_{\text{crit}} (d\rho/d\ln f)$  
and $\lambda/mm=M^{-1}_{d}\leq 1/R$ with $M=1$TeV (a weak coupling according to gravitational Table-Top experiments \cite{Carena:2001xy}).} 
\label{evolution_deSitter}
\end{figure*}

Through a numerical analysis of Eq. \eqref{eq:desitter brane}, we have Figure \ref{evolution_deSitter}, showing that these solutions are appropriate description from a source where GWs have been emitted in comparison to the Minkowski case, the free parameters in the numerical solution are selected through Ref. \cite{Lin:2020wnp}.  The fluctuations considered provide a consistent truncation of the fluctuated RS-dimensional theory, meaning that our numerical solutions will also be linear solutions to a D-dimensional theory with all metric fluctuations.

\section{Using GW170817 and S190425z as standard sirens for the braneworld}
\label{Sec:Limits}

Using the energy limits on GW propagation of the brane model, the next question that arise is that if we can distinguish these cases from the standard concordance model. Therefore, to obtain an answer, we compare our exact solutions in both energy limits on the GW propagation that already has be obtained by the standard sirens provided by the detection of the GWs from the neutron stars binary GW170817 and from S190425z detection. Then, we select the modifications of the perturbation equations over the cosmological background. From the identification of the electromagnetic counterpart the redshift of the source can be constrained by $z=0.01$. In both cases we are at very small redshifts. Since we have two concrete cases, we can specify the luminosity distances for both components and consider a Bayesian estimation for the $\Lambda$CDM model using CMB+BAO+SNeIa samplers. Other effect that should be taken into account is the Shapiro gravitational delay which affects our results from GWs. For this reason, we present in Appendix \ref{Sec. Shapiro}, a short revision of this effect discussing the consequences in our results.

\begin{figure*}
\centering
\includegraphics[width=0.5\textwidth,origin=c,angle=0]{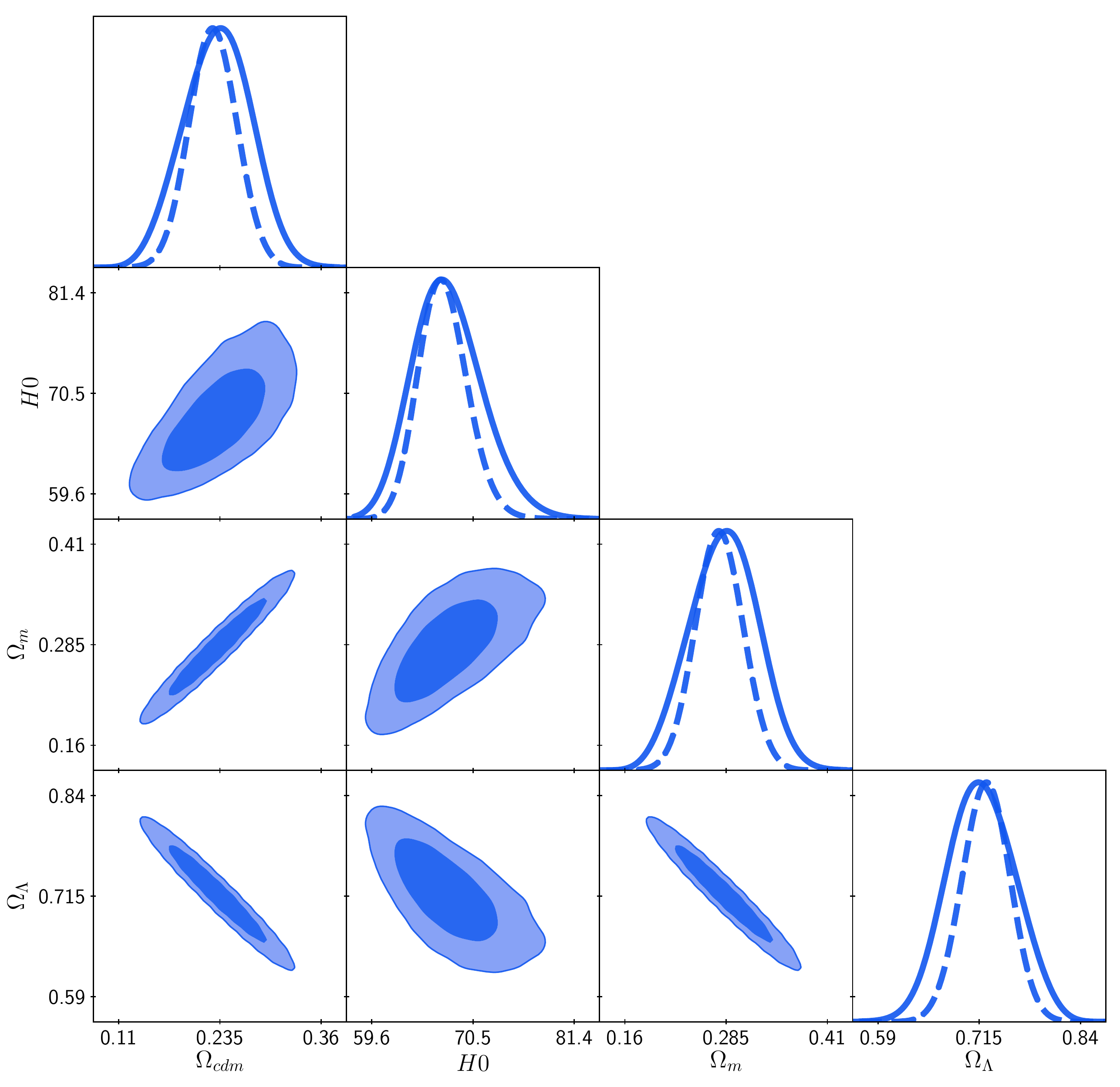}
\includegraphics[width=0.4\textwidth,origin=c,angle=0]{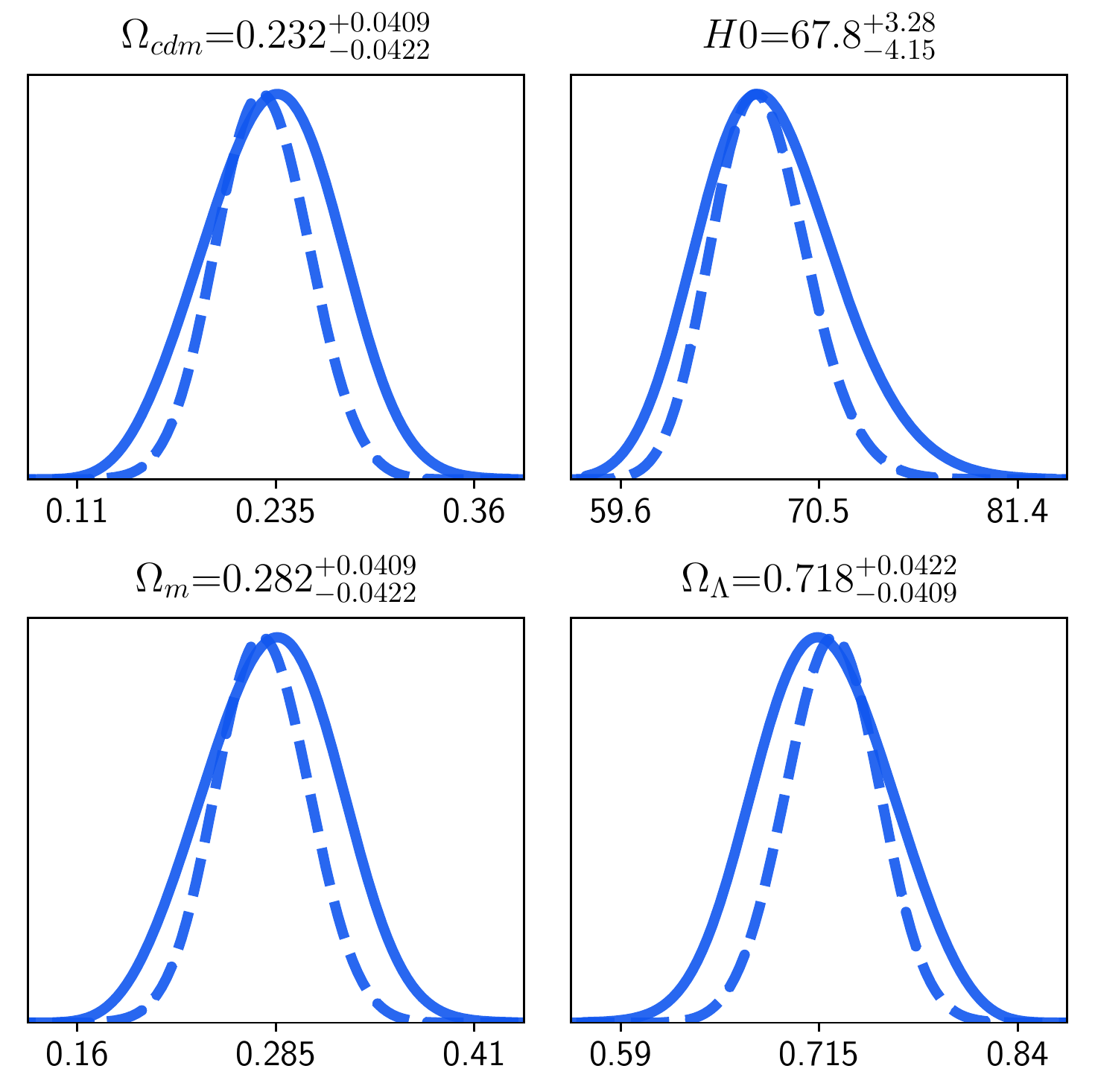}
\caption{Joint posterior distribution for the Minkowski brane case using GW170817 and S190425z detections. The dark blue (light blue) regions correspond to 68\% (95\%) C.L. contours, respectively.} 
\label{posterior_Minkowski}
\end{figure*}

\begin{figure*}
\centering
\includegraphics[width=0.5\textwidth,origin=c,angle=0]{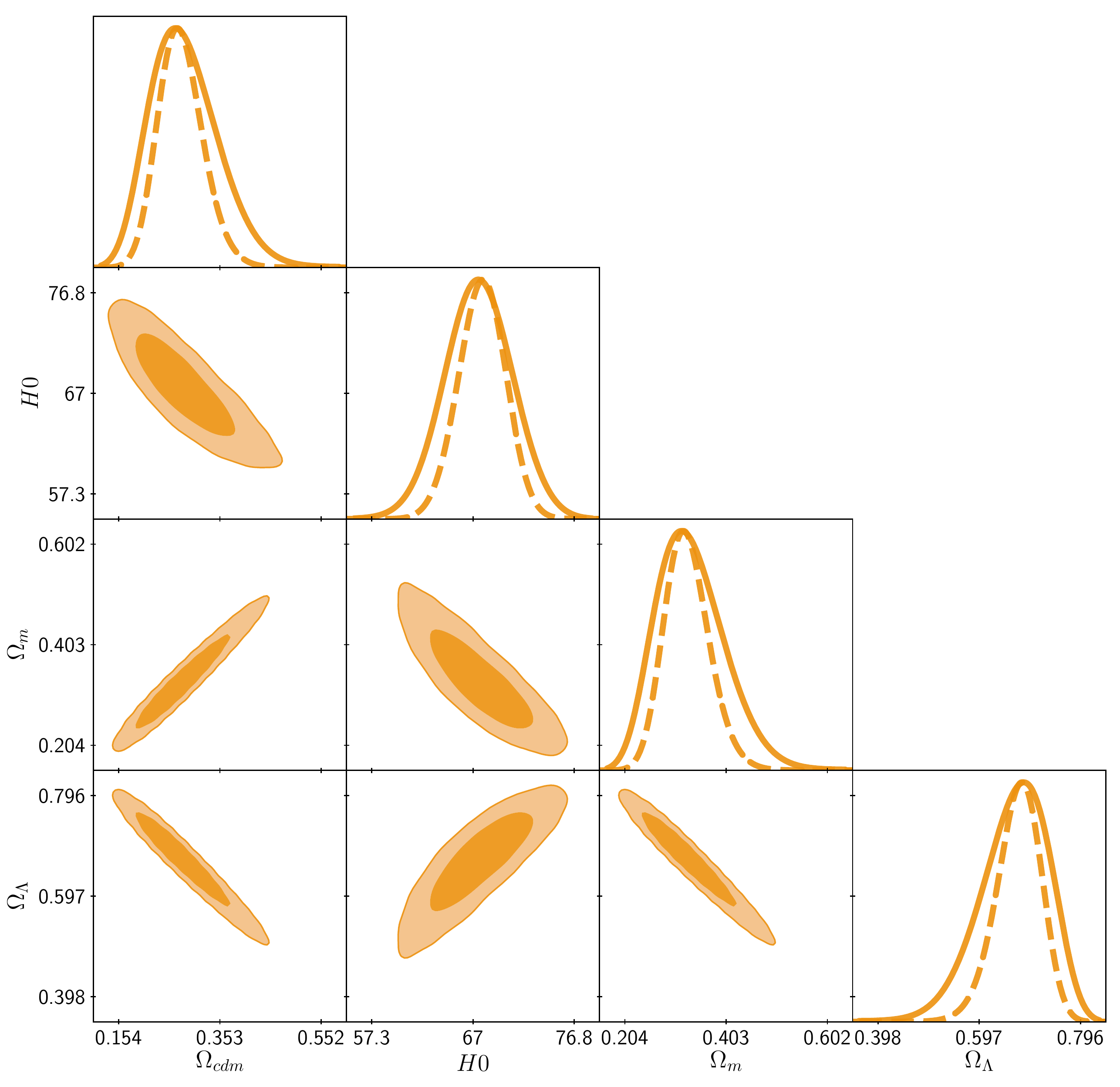}
\includegraphics[width=0.4\textwidth,origin=c,angle=0]{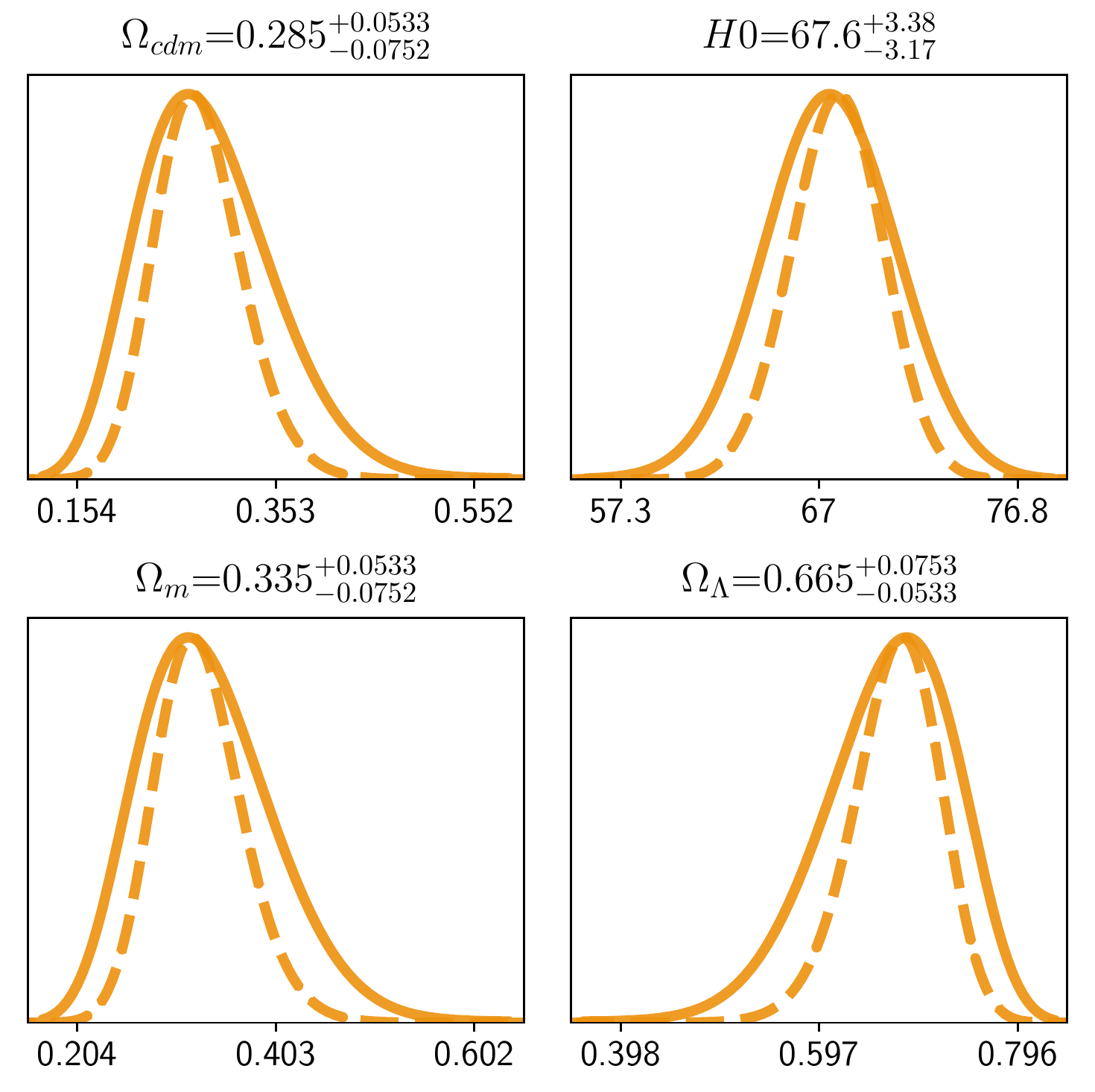}
\caption{Joint posterior distribution for the De Sitter brane case using GW170817 and S190425z detections. The dark orange (light orange) regions correspond to 68\% (95\%) C.L. contours, respectively.} 
\label{posterior_deSitter}
\end{figure*}

To perform Bayesian analyses it is useful to understand in physical terms with who compare these cases using the standard $\Lambda$CDM as a pivot model.
For this goal, we start from a generic $w$CDM model, with a fixed value of $w_0$, and set the measurements of the luminosity distances with standard sirens that could help in discriminating it from $\Lambda$CDM.
Let us consider
\begin{eqnarray}
&&d_L(z;H_0,\Omega_m,w_0)=\frac{1+z}{H_0}\times\nonumber\\&&\int_0^z \frac{d\tilde{z}}{\sqrt{ \Omega_m (1+\tilde{z})^3 +(1-\Omega_m) (1+\tilde{z})^{3(1+w_0)}}   },
\end{eqnarray}
where we have written explicitly the dependence of $d_L$ on the cosmological parameters and we consider them with the samplers mentioned using: $H_0 =67.64$(Km/s)/Mpc and $\Omega_m =0.309$, as fiducial values for $\Lambda$CDM.  After this, we can compute the quantity
\begin{equation}\label{eq:dl}
\frac{\Delta d_L}{d_L}\equiv \frac{d_L(z;H_0,\Omega_m,w_0)-d^{\Lambda\rm CDM}_L(z;H_0,\Omega_m)}{d^{\Lambda\rm CDM}_L(z;H_0,\Omega_m)}\, .
\end{equation}
This represent the difference between the luminosity distance in $w$CDM with a given value of $w_0$, and the luminosity distance in $\Lambda$CDM (where $w_0=-1$), at fixed $\Omega_m,H_0$. Similarly, we can compute this quantity for each brane limit model by considering $d_L$ in Eq.(\ref{eq:dl}) of the pivot model now as $d^{gw}_{L}$ as in Eq.(\ref{eq:dl_gw}). 
We construct the 2D joint contours plots of the cosmological parameters in comparison to $\Lambda$CDM model with both GW detections with a Markov Chain Monte Carlo (MCMC) simulation. The best fits for the cosmological variables are reported in Table \ref{tab:Minkowski brane}. For our exact solutions, we have the posteriors respectively showed in Figures \ref{posterior_Minkowski} and \ref{posterior_deSitter}. We notice that the $H_0$ tension seems to be relaxed at least $0.2\%$ in the Minkowski case in comparison to the De Sitter case. The region contour plot, $\Omega_m-\Omega_\Lambda$ is the primary constraint on dark energy. As a geometry probe  GW detections drive most of the statistical information and the constraint is much better in the De Sitter case. Interesting enough, these results can drive an even more drop of the tension if we consider future GRB and WL detections.

\begin{table*}
\caption{Parameters and mean values for the Minkowski and De Sitter brane case using GW170817 and S190425z detections.}
\centering
\begin{tabular}{|ccccc|}
\hline
Parameter     &  \quad  Best-fit     & \quad   mean$\pm\sigma$ &\quad $95\%$ lower & \quad $95\%$ upper \\
\hline
\multicolumn{5}{|c|}{Minkowski case} \\ [0.4ex]
\hline
$\Omega_{cdm}$  & $0.2213$ & $0.2316^{+0.041}_{-0.042}$ &  $0.1503$ & $0.3137$  \\ [0.4ex]
$H_0$  & $66.78$ & $67.81^{+3.3}_{-4.2}$ &  $60.48$ & $75.63$ \\ [0.4ex]
$\Omega_m$  & $0.2713$ & $0.2816^{+0.041}_{-0.042}$ &  $0.2003$ & $0.3637$ \\ [0.4ex]
$\Omega_{\Lambda}$  & $0.7286$ & $0.7183^{+0.042}_{-0.041}$ &  $0.6363$ & $0.7996$ \\ [0.4ex]
\hline
\multicolumn{5}{|c|}{De Sitter case} \\ [0.4ex]
\hline 
$\Omega_{cdm}$  & $0.2699$ & $0.2851^{+0.053}_{-0.075}$ &  $0.1578$ & $0.4227$  \\ [0.4ex]
$H_0$  & $68.11$ & $67.6^{+3.4}_{-3.2}$ &  $60.98$ & $74.13$ \\ [0.4ex]
$\Omega_m$  & $0.3199$ & $0.3351^{+0.053}_{-0.075}$ &  $0.2078$ & $0.4727$ \\ [0.4ex]
$\Omega_{\Lambda}$  & $0.68$ & $0.6648^{+0.075}_{-0.053}$ &  $0.5272$ & $0.7921$ \\ [0.4ex]
\hline
\end{tabular}
\label{tab:Minkowski brane}
\end{table*}

\section{Discussion and Conclusions} 
\label{DC}

In this paper we study a brane world theory based in RS model in order to analyze if the theory is consistent with the constraints obtained from GW170817 and S190425z events. Through the weak field limit for braneworlds, our results shown that in this scenario the speed propagation of GW is equal to $c$, a results that is in agreement with GR. However, the extra components that comes from five dimensional physics in Eqs. \eqref{1GW}-\eqref{01GW}, introduce a Weyl tensor that plays an important role. From Eq. \eqref{01GW} we notice that the brane tension is an essential ingredient and represents the difference with the standard knowledge of a GW in GR, indeed, using previous constraints we found the most restrictive as $\sim1.03\times10^6$eV$^{-2}$ for Table Top experiments that set a cutoff over the value of $\lambda$. Even more, we calculate the cosmic tensor perturbations in this scenario, dividing our studies in Minkowski (low energies) and De Sitter (higher energies) cases, presenting the evolution of the GW in conformal time. The Minkowski case presents a behavior that mimic a damped oscillator while De Sitter case presents a subtle increase of oscillations due the topology involved in this case.
As a final study, we perform statistical studies from standard sirens definitions and obtaining a relaxation of the tension presented in the values of $H_0$ of at least $0.2\%$ in the Minkowski case in comparison to the De Sitter case. This result is an incentive to keep exploring theories that involves extra-dimensions and wait for more data in order to increase the statistics involved.

\section*{Acknowledgements}
M.A.G.-A. acknowledges support from SNI-M\'exico, CONACyT research fellow, COZCyT, Instituto Avanzado de Cosmolog\'ia (IAC) collaborations and CONICYT REDES (190147).
CE-R acknowledges the Royal Astronomical Society as FRAS 10147 and supported by PAPIIT Project IA100220 and ICN-UNAM projects. Also, would like to acknowledge the contribution of the COST Action CA18108. The authors thank the anonymous reviewers whose comments have improved this manuscript and also acknowledge conversations with L. Arturo Ure\~na. 
\\

\appendix

\section{Other cases in the weak field limit} 
\label{Appendix}

In order to complement the results for the gravitational wave equation, we present the other cases for the subscripts $\mu$, $\nu$, $\beta$ and $\delta$. Hence, for $\mu,\nu=2,2$ we have

\begin{eqnarray}
&&\Box h_{22}-(\eta_{2[2}\Box h_{\delta]\beta}-\eta_{\beta[2}\Box h_{\delta]2}-2h_{2\delta,\beta2})n^{\beta}n^{\delta}\nonumber\\&&+\mathcal{O}(5D)=0,
\end{eqnarray}

\begin{itemize}
\item When $\beta=\delta=2$
\begin{eqnarray}
&& \Box h_{22}+2h_{22,22}n^2n^2+\mathcal{O}(5D)=0.
\end{eqnarray}

\item When $\beta=2$, $\delta=3$
\begin{eqnarray}
&& \Box h_{22}+2h_{23,22}n^2n^3+\mathcal{O}(5D)=0.
\end{eqnarray}

\item When $\beta=3$ y $\delta=2$
\begin{eqnarray}
&&\Box h_{23}+2h_{22,32}n^{3}n^{2}+\mathcal{O}(5D)=0.
\end{eqnarray}

\end{itemize}

Hence, for $\mu=2$ and $\nu=3$, we have

\begin{eqnarray}
&&\Box h_{23}-(\eta_{2[3}\Box h_{\delta]\beta}-\eta_{\beta[3}\Box h_{\delta]2}-2h_{2\delta,\beta3})n^{\beta}n^{\delta}\nonumber\\&&+\mathcal{O}(5D)=0, 
\end{eqnarray}

\begin{itemize}

\item When $\beta=2=\delta$
\begin{eqnarray}
&&\Box h_{23}+2h_{22,23}n^{2}n^{2}+\mathcal{O}(5D)=0.
\end{eqnarray}

\item When $\beta=2$ y $\delta=3$
\begin{eqnarray}
&&\Box h_{23}+2h_{23,23}n^{2}n^{2}+\mathcal{O}(5D)=0.
\end{eqnarray}

\item When $\beta=3$ y $\delta=2$
\begin{eqnarray}
&&\Box h_{23}+(\Box h_{22}+\Box h_{33}-2h_{22,33})n^{3}n^{2}\nonumber\\&&+\mathcal{O}(5D)=0.
\end{eqnarray}
\end{itemize}

\section{Shapiro delay new constrains} 
\label{Sec. Shapiro}

To extend the physics behind the constrains found, we extend the description of the the Shapiro delay in our scenario. First, we write the proper time on the bulk for a null geodesic ($ds^2=0$) in the form \cite{Visinelli:2017bny}
\begin{equation}
    d\tau^2=V(r)^{-2}dr^2+V(r)^{-1}r^2d\Omega^2+V(r)^{-1}f(R)^{-2}dR^2,
\end{equation}
where $V(r)=1-2M/r$, being $M$ a near point mass in Schwa-rzschild coordinates, $f(R)=(R/\ell)^2$ is related with the AdS$_5$ scale factor and $\ell$ is the curvature radius of the AdS$_5$. The first two terms corresponds to the standard Shapiro delay well known in four-dimensional physics, while the last terms is the correction provided by five dimensional physics. Notice how the curvature radius is one of the main parameters that give us information about the presence of corrections by extra-dimensions. In this sense in \cite{Visinelli:2017bny} was obtained  a constraint of the order $\ell < 0.535$ Mpc, being not competitive in comparison with the constraint produced by black holes X-ray binaries as $\ell\lesssim10^{-2}$mm \cite{Emparan:2002jp}, stressing that corrections produced in this scenario the Shapiro effect can be considered as negligible. 


\bibliographystyle{spphys}
\bibliography{librero1.bib}

\end{document}